\documentclass[12pt]{article}
\usepackage{amsmath}
\usepackage{amssymb}
\usepackage{graphicx}
\usepackage{enumerate}
\usepackage{float}
\usepackage{natbib}
\usepackage{url} 
\newcommand{\blind}{1}
\usepackage[font=small,skip=0pt]{caption}
\usepackage[ruled]{algorithm2e}
\usepackage{hyperref}

\usepackage{array}
\usepackage{multirow}

\begin{document}

\def\spacingset#1{\renewcommand{\baselinestretch}%
{#1}\small\normalsize} \spacingset{1}

\if1\blind
{
  \title{\bf Distributed Kaplan-Meier Analysis via the Influence Function with Application to COVID-19 and COVID-19 Vaccine Adverse Events}
  \author{Malcolm Risk\hspace{.2cm}\\
    Department of Biostatistics, University of Michigan\\
     Xu Shi \\
    Department of Biostatistics, University of Michigan \\
    Lili Zhao \\
    Department of Preventive Medicine, Northwestern University}
  \maketitle
} \fi

\if0\blind
{
  \bigskip
  \bigskip
  \bigskip
  \begin{center}
    {\LARGE\bf Distributed Kaplan-Meier Analysis via the Influence Function with Application to COVID-19 and COVID-19 Vaccine Adverse Events}
\end{center}
  \medskip
} \fi

\bigskip
\begin{abstract}
During the COVID-19 pandemic, regulatory decision-making was hampered by a lack of timely and high-quality data on rare outcomes. Studying rare outcomes following infection and vaccination requires conducting multi-center observational studies, where sharing individual-level data is a privacy concern. In this paper, we conduct a multi-center observational study of thromboembolic events following COVID-19 and COVID-19 vaccination without sharing individual-level data. We accomplish this by developing a novel distributed learning method for constructing Kaplan-Meier (KM) curves and inverse propensity weighted KM curves with statistical inference. We sequentially update curves site-by-site using the KM influence function, which is a measure of the direction in which an observation should shift our estimate and so can be used to incorporate new observations without access to previous data. We show in simulations that our distributed estimator is unbiased and achieves equal efficiency to the combined data estimator. Applying our method to Beaumont Health, Spectrum Health, and Michigan Medicine data, we find a much higher covariate-adjusted incidence of blood clots after SARS-CoV-2 infection (3.13\%, 95\% CI: [2.93, 3.35]) compared to first COVID-19 vaccine (0.08\%, 95\% CI: [0.08, 0.09]). This suggests that the protection vaccines provide against COVID-19-related clots outweighs the risk of vaccine-related adverse events, and shows the potential of distributed survival analysis to provide actionable evidence for time-sensitive decision making. 
\end{abstract}

\noindent%
\vfill

\newpage
\spacingset{1.9} 
\section{Introduction}
\label{sec:intro}

\subsection{Rare Adverse Events following COVID-19 Vaccination}

On April 13th, 2021 the FDA and CDC jointly announced a pause on the Johnson \& Johnson COVID-19 vaccine due to six reports of rare and severe blood clots following vaccination. This pause remained in effect until it was lifted on April 23rd, 2021 when the FDA and CDC concluded that such events were sufficiently rare for the benefits of vaccination to outweigh any known and potential risks \citep{CDC2021}. Although this process took only ten days, this was at a critical moment in the vaccine rollout and occurred when overall supply of vaccines was still limited. Timely and reliable information regarding the incidence of these rare adverse events could have allowed officials to avoid or shorten this pause, but the data available to FDA and CDC officials were inadequate. 

The Johnson \& Johnson pause is a particular example of a much wider problem, where timely information on rare adverse events from medical treatments cannot come from phase 3 trials (due to insufficient sample size) or self-reported data (due to poor data quality). Contemporaneous data on the adverse events in the US from voluntary reports through the Vaccine Adverse Events Reporting system (VAERS) are subject to inaccuracy, lack of verifiability, and spurious association due to lack of access to a control group of untreated individuals \citep{Shimabukuro2015, VAERS2025}. To provide better evidence, we need to accelerate the pace of research that leverages large quantities of routinely collected electronic health record (EHR) data. In contrast to self-reported data, EHR data are less affected by reporting bias and allow investigators access to an untreated comparator group. Amassing the high sample size of EHR data required to study rare adverse events has motivated the establishment of federated networks combining data from multiple centers, such as the CDC's Vaccine Safety Datalink \citep{VSD2025}. However, sharing individual-level data is subject to lengthy approval processes to protect patient privacy, limiting the creation, expansion, and usage of such networks. To avoid sharing individual-level data, we need to develop distributed methods for conducting multi-center analyses.

\subsection{Scientific Goal: Multi-center Study of Adverse Events}

The scientific goal of this paper is to use data from three large healthcare centers to conduct an observational study of thromboembolic events (blood clots) following COVID-19 vaccination and SARS-CoV-2 infection, without sharing any individual-level data between sites. Specifically, we address two scientific questions: 

\begin{enumerate}
\item	How does the risk of thromboembolic events following first COVID-19 vaccine dose compare to the risk of thromboembolic events for unvaccinated individuals with a SARS-CoV-2 infection?
\item	Does vaccination with BNT162b2 (Pfizer) or mRNA-1273 (Moderna) result in a higher risk of thromboembolic events?
\end{enumerate}

Given our data sharing restrictions and clinical setting, we needed a method with the following characteristics:

\begin{enumerate}
\item Ability to account for censoring and loss to follow-up in time-to-event data.
\item No requirement to share individual-level data.
\item No restrictive proportional hazards assumption given the likelihood of adverse events clustering near the beginning of the follow-up. 
\item Ability to incorporate data from sites with very low event counts due to the rare nature of vaccine adverse events. 
\item Adjustment for confounding in observational data.
\end{enumerate}

As no existing method met all of these requirements (see next section), we developed a novel method for estimating Kaplan-Meier (KM) \citep{KM1958} and inverse propensity weighted Kaplan-Meier (IPW-KM) curves from multiple sites with no sharing of individual-level data to assist in answering our scientific question. As far as we know, this is the distributed method developed for constructing Kaplan-Meier curves. The choice of KM curves as an analysis method is motivated by the structure of our data; we want to study time from vaccination to adverse event in the presence of censoring due to subsequent vaccine dose, infection, or death. The use of IPW-KM curves is crucial to address potential confounding, particularly with regards to our first research question, where we would expect substantial confounding due to vaccine recipients having more comorbidities and skewing much older than unvaccinated individuals. 

\subsection{Distributed Survival Analysis: Existing Methods}

Existing work on distributed survival analysis \citep{Duan2020, Lu2015, Li2022, Wu2021} focuses on parametric models and the semi-parametric Cox proportional hazards model. The Cox model is not appropriate for our use case as we would expect strong violation of the proportional hazards assumption, with adverse events clustering near the beginning of the follow-up after vaccination or infection. Fitting a more flexible Cox model or selecting the correct structure for a parametric model would require repeated assessment of assumptions and multiple runs of model fitting, which is a difficult task in a distributed setting. Hence we chose to use KM and IPW-KM curves to make the fewest assumptions possible.

There is an existing school of work on distributed KM curves that relies on either multiparty encryption protocols or the addition of differentially private noise to survival times to protect patient privacy \citep{Spath2022, Froelicher2022, Vogelsang2020, Bonomi2020}. Both methods still require sharing of information in a similar structure and dimension to the original dataset and so may need to meet a high bar for data sharing approval in practice. They also require an unrealistic degree of coordination and trust from investigators at individual sites who are unlikely to be familiar or comfortable with complex, non-trivial  methods. For our specific use case, neither method met our requirement for easy implementation, low dimension of shared data, and comprehensive methods for statistical inference and confounder adjustment.

\subsection{Distributed Survival Analysis: Our Approach}

In this paper, we present a comprehensive distributed learning method for estimating KM curves without sharing any patient-level data. Our method offers multiple practical benefits compared to existing work:
\begin{enumerate}
\item	No requirement to share individual-level data of any form (including survival times), avoiding privacy concerns and a lengthy approval process.
\item	Confounder adjustment using inverse propensity weighting (IPW), with accompanying statistical inference and hypothesis testing using a distributed log-rank test.
\item	Informative data visualization of the differences in survival probability between groups.
\end{enumerate}
Our method is sequential rather than centralized, meaning that each site shares updated summary statistics with only the next site in a prespecified order. We start by constructing an estimated curve that can be shared from one site to another. Based on the influence function of the KM estimator \citep{Reid1981}, which is a measure of the direction in which an observation should affect our estimate, we update that curve at each subsequent site before passing the new estimates onward. We extend our method to adjust for confounders by estimating IPW KM curves \citep{Xie2005}, using existing renewable estimation methods to fit the propensity score model \citep{Luo2020}. To accomplish our goal, we make two important methodological contributions: \begin{enumerate}
\item	A novel updating equation to incorporate new data into a survival probability estimate, allowing for sites as small as one observation with no events. 
\item	A novel distributed approach for statistical inference based on the influence function that allows us to construct asymptotic 95\% confidence intervals for survival probabilities and a two-sample distributed log-rank test in KM and IPW-KM curves. 
\end{enumerate}
In Section 2 (Methods) of this paper we show how to update a KM curve site-by-site. We show how to derive 95\% confidence intervals and a log-rank test at the final site, and extend our method to the IPW case. In Section 3 (Simulations), we show that our method is unbiased and has minimal loss of efficiency compared to KM analysis in the pooled data.
In Section 4 (Application), we exploit this new methodology to compare incidence of blood clots between mRNA-vaccinated and SARS-CoV-2 infected individuals with data from three large sites, demonstrating the utility of our method for studying rare events by combining massive data across multiple sites. We provide an easy-to-use R package  and give guidance on parameter choices, with extensive detail given in the supplementary material and package documentation. We discuss the study results and potential extensions in Section 5 (Conclusion).

\section{Methods}
\label{sec:meth}

\subsection{Notation}

In survival analysis, each subject $i$ is considered to have a true survival time $T_i$ representing time-to-death or disease onset and censoring time $C_i$ representing time to which that subject was observable. In a survival setting we do not observe true survival or censoring time for individual $i$, but instead follow-up time $X_i = \text{min}(T_i,C_i)$ and censoring indicator $\Delta_i=I(T_i<C_i)$. 

\subsection{Kaplan-Meier Estimator}

Suppose that we have N total subjects with follow-up times $\mathbf{X} = (X_1,...,X_N)$ and censoring indicators $\mathbf{\Delta} = (\Delta_1,...,\Delta_N)$. We will assume these subjects are divided across K sites with $n_k$ observations each, such that $\sum_{k=1}^K n_k= N$. We assume the existence of a continuous and differentiable true survival function $S(t) = P(T_i>t)$, representing the probability of being alive (or disease-free, depending on the outcome chosen) prior to t. In a single-center study, we could obtain the KM estimate for $S(t)$ as:
\begin{equation}\label{eq:KM}
\hat{S}(t; \mathbf{X},\mathbf{\Delta})=\prod_{j:t_j\leq t} 1- \frac{d_j}{n_j}
\end{equation}
where $d_j = \sum_{i=1}^N \Delta_i I(X_i=t_j)$ is the observed number of events at $t_j$, $n_j =\sum_{i=1}^NI(X_i>t_j)$ the number of individuals at-risk at $t_j$,  and $\{t_j:t_j \leq t\} = \{X_i \in \mathbf{X} : X_i \leq t,\Delta_i=1\}$ is the set of unique event times observed prior to time of interest $t$ in ascending order. The numerator of the fraction in equation \eqref{eq:KM} is the number of individuals with an event at $t_j$ and the denominator is the number of individuals in the risk set (i.e., event-free and uncensored) past $t_j$. In a multi-center study, equation \eqref{eq:KM} can be calculated only if we pool all the individual-level survival data together. Our goal in this paper was to avoid any such pooling, while preserving the non-parametric benefits of KM estimation, namely the ability to approximate an arbitrarily complex curve with an increasing degree of flexibility as sample size increases. This is a more complex task than many other distributed learning methods because the growing complexity of the curve across sites makes it more difficult to characterize the curve using a fixed set of summary statistics.

\subsection{Our Approach: Online Updating via the Influence Function}

\subsubsection{The Updating Equation}

In 1979, the computer scientist D.H.D. West published an algorithm for avoiding approximation errors when calculating sample mean and variance \citep{West1979}. The algorithm, now widely used, follows a simple online updating approach to calculating the sample mean:
\begin{enumerate}
\item	Set $\overline{X}_1=X_1$.
\item   For $i=2,\dots,N$ set $\overline{X}_i = \overline{X}_{i-1} + \frac{X_i - \overline{X}_{i-1}}{i}$.
\end{enumerate}
We note two important features of this algorithm. Firstly, updating step (2) only requires access to the current running mean and not previous observations. Secondly, each observation is incorporated by adding a residual $X_i-\overline{X}$, weighted by the cumulative sample size. 

A West-type algorithm would be most researchers' first instinct when calculating simple estimates such as proportions and sample variances in a distributed or online context. Currently, West-type algorithms are not available for more complex estimators (such as KM curves and regression coefficients), because there is no clear concept of a ``residual" associated with each parameter estimate. 

In this paper, we describe a much more general set of online algorithms by considering the ``residual" in West-type algorithms as an estimate for the \textit{influence function}. Given an underlying model $\mathbb{P}$ and functional $\Psi$ (e.g. mean, covariance) an \textit{influence function} is the derivative of a target parameter $\Psi(\mathbb{P})$ with respect to a perturbation in the direction of a submodel defined by the Dirac measure of a single observation: 
\begin{equation}\label{eq:IF}
\psi(X) = \frac{\partial\Psi(\mathbb{P}_{\epsilon})}{\partial \epsilon},
\end{equation}
where $X$ is an observation and $\mathbb{P}_\epsilon(x) = (1-\epsilon)\mathbb{P}(x) + \epsilon \delta_X$ for $\delta_X$ the Dirac measure at $x=X$ \citep{kennedy2023}. In West's algorithm, the residual $X_i-\overline{X}$ can be interpreted as an estimate for the influence function of the sample mean, $\psi(X_i)=X_i-\mu$ with $\overline{X}$ replacing $\mu$. The key novel idea in our paper is that since the influence function is a measure of the direction in which a single observation would shift our estimate, it can be used to update an existing estimate without requiring access to any other observations. 

Returning to our application, we now show how this idea can be used to construct an updating equation for KM estimates. As derived by \cite{Reid1981}, the influence function of the KM estimator is given by:
\begin{equation}\label{eq:IFKM_Reid}
\psi(X_i,\Delta_i)(t)= -S(t)\left[\frac{\Delta_i I(X_i\leq t)}{Y(X_i)} -\int_0^{\text{min}(X_i, t)} \frac{\lambda(u)}{Y(u)}du \right]         \end{equation}
where $Y(t)=P(X_i>t)$ is the at-risk function and $\lambda(t)=-\frac{S^{'}(t)}{S(t)}$ the hazard function. We go into greater detail regarding the influence function and its derivation in Appendix A. 

More rigorously, we express the relationship between the influence function and the KM estimate for $N$ observations in terms of a Von Mises expansion, which is the equivalent of the Taylor series expansion for functionals \citep{Gill1989}:
\begin{equation}\label{eq:VonMises}
\begin{split}
\hat{S}(t, \mathbf{X}, \mathbf{\Delta}) = S(t) + \frac{1}{N}\sum_{i=1}^N\psi(S;X_i, \Delta_i)(t) + o_p\left(\frac{1}{N^{1/2}}\right).
\end{split}
\end{equation}
Now consider two KM estimates: $\hat{S}_N$, incorporating $N$ observations, and $\hat{S}_{N+1}$, incorporating one additional observation. After some simple algebra, we can obtain an expression for $\hat{S}_{N+1}$ in terms of $\hat{S}_N$ by subtracting their Von Mises expansions:
\begin{equation}\label{eq:AsympUpdate}
\begin{split}
\hat{S}_{N+1}(t) = \hat{S}_N(t) + \frac{\psi(S;X_{N+1}, \Delta_{N+1})(t)}{N+1} + o_p\left(\frac{1}{N^{1/2}}\right)
\end{split}
\end{equation}
Providing that we can calculate an estimate $\hat{\psi}$ for the influence function, we now have a natural updating formula for the KM estimator at a particular time $t$:
\begin{equation}\label{eq:Update}
\begin{split}
\hat{S}_{N+1}(t) = \hat{S}_N(t) + \frac{\hat{\psi}(\hat{S}_N;X_{N+1}, \Delta_{N+1})(t)}{N+1}
\end{split}
\end{equation}
The derivation in equations \eqref{eq:VonMises}-\eqref{eq:Update} is not specific to the KM estimator, and can be applied to any estimator with a valid Von Mises expansion. For example, replacing $S_N$ by $\overline{X}_N$ and the influence function $\hat{\psi}$ by $X_{N+1} - \overline{X}_{N}$ would give us West's algorithm. In a distributed learning setting, this means that each site can calculate the influence function value for each observation and use those values to shift an estimate received from the previous site. Then the new estimates can be passed on to the next site, updated, and passed further along until all observations have been incorporated. 

For our application, these equations allow us to update an existing estimate of survival probability based on a single new observation, without needing access to any individual-level data from previous sites. This is a critical innovation because in multi-center studies where data are sparsely spread across many sites, we can incorporate data from sites with as low as just one observation (censored or uncensored). We frequently observe such data when dealing with rare events and exposures.

\subsubsection{Algorithmic Implementation}

\begin{figure}
\begin{center}
\includegraphics[width=6in]{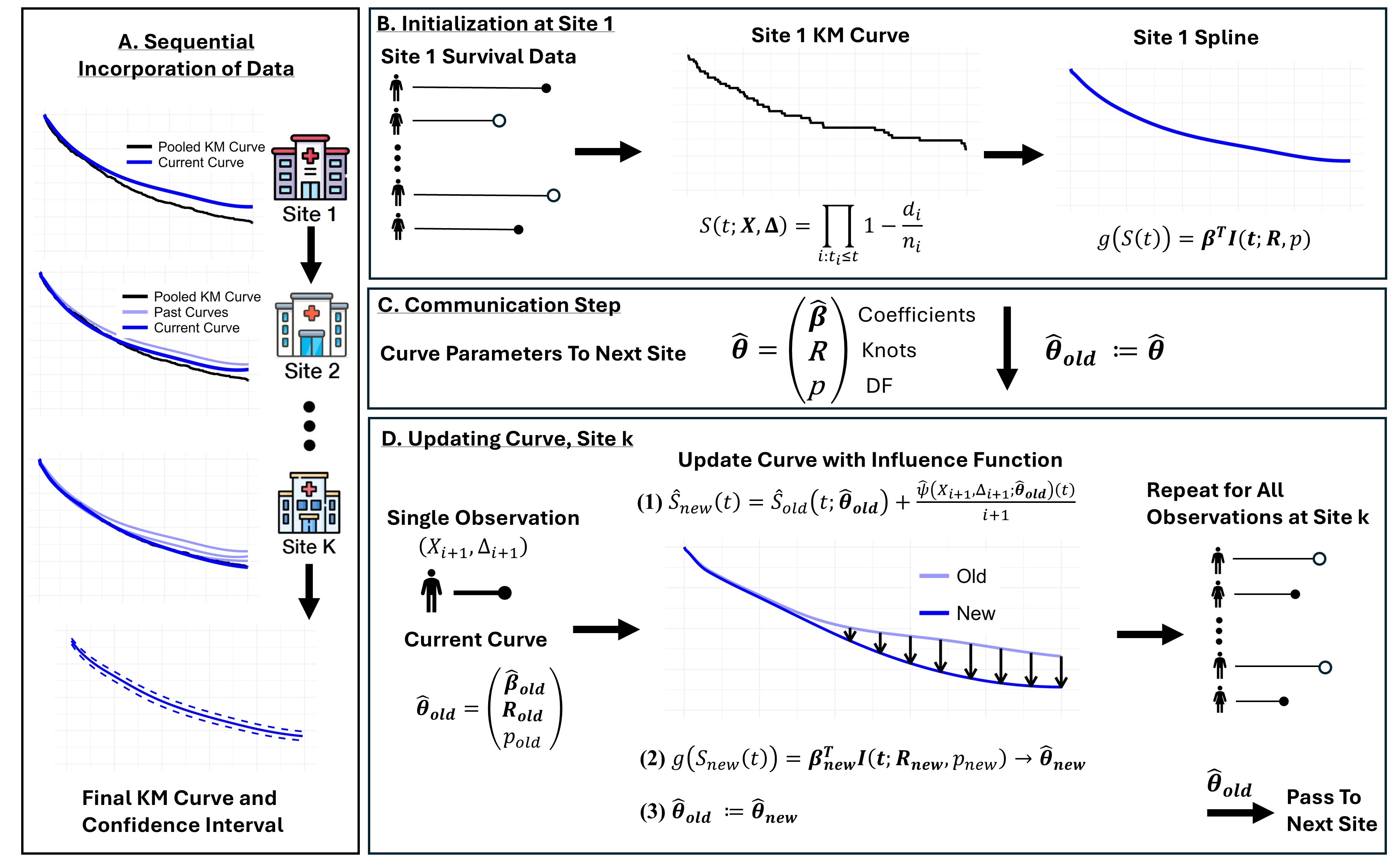}
\end{center}
\caption{Schematic of update steps in distributed KM method. \label{fig:first}}
\end{figure}

The updating equation \eqref{eq:Update} pertains only to individual time points, while in practice we often want to estimate the entirety of the survival curve. In addition, updating the survival curve $\hat{S}$ using the influence function \eqref{eq:IFKM_Reid} requires an estimate for the nuisance at-risk curve $\hat{Y}$. For this purpose, we use a continuous spline function characterized by parameters $\hat{\boldsymbol{\theta}}=(\hat{\boldsymbol{\alpha}}, \hat{\boldsymbol{\beta}}, \textbf{R}, p)$ where $\hat{\boldsymbol{\alpha}}$ and $\hat{\boldsymbol{\beta}}$ are coefficients for survival and at-risk curves, $\textbf{R}$ are knot locations, and $p$ is the number of spline degrees of freedom. Figure \ref{fig:first} gives a visual illustration of how our method updates the estimated curve across sites. 

Specifically, we obtain a spline approximation based on the $n_1$ observations at the first site by estimating a KM curve and then using the survival probability estimates $\hat{S}(t)$ in a spline regression model to obtain $\hat{\boldsymbol{\beta}}_{n_1}$:
\begin{equation}\label{eq:SplineReg}
g(\hat{S}(t))=\boldsymbol{\beta}^T\textbf{I}(t;\textbf{R}, p)
\end{equation}
where $\textbf{I}$ is a differentiable spline basis function and $g$ a given link function (Figure \ref{fig:first} Panel B and C). We fit a similar model to obtain $\hat{\boldsymbol{\alpha}}_{n_1}$ and then pass the full set of parameters $\hat{\boldsymbol{\theta}}_{n_1}$ to the second site, where we then update the estimate observation-by-observation. 

Consider the update step for the first observation at the second site ($X_{n_1+1}, \Delta_{n_1+1}$). Our first step is to obtain an estimate of the influence function of our observation based on our existing parameters. The three key functions $Y$, $S$, and $\lambda$ are estimated from the spline parameters as follows:
\begin{equation}\label{eq:IF_Params}
\begin{split}
\hat{S}(t; \hat{\boldsymbol{\theta}}_{n_1})&=g^{-1}\left(\hat{\boldsymbol{\beta}}_{n_1}^T\textbf{I}(t;\textbf{R}_{n_1}, p_{n_1})\right)\\
\hat{Y}(t; \hat{\boldsymbol{\theta}}_{n_1})&=g^{-1}\left(\hat{\boldsymbol{\alpha}}_{n_1}^T\textbf{I}(t;\textbf{R}_{n_1}, p_{n_1})\right) \\
\hat{\lambda}(t;\hat{\boldsymbol{\theta}}_{n_1})&= -\frac{\hat{S}(t; \hat{\boldsymbol{\theta}}_{n_1})}{\partial\hat{S}(t; \hat{\boldsymbol{\theta}}_{n_1})/\partial t}
\end{split}
\end{equation}
We index the spline knots and degrees of freedom $\textbf{R}_{n_1}$ and $p_{n_1}$ by sample size because our algorithm allows for changing the knot locations and degrees of freedom as new observations are incorporated. To estimate the influence function, we plug the estimates from equation \eqref{eq:IF_Params} into equation \eqref{eq:IFKM_Reid} and use numerical integration techniques to obtain $\hat{\psi}(X_{n_1+1}, \Delta_{n_1+1};\hat{\boldsymbol{\theta}}_{n_1})$. We then obtain shifted curve estimates at a discrete set of time points using equation \eqref{eq:Update} and obtain the new estimate $\hat{\boldsymbol{\beta}}_{n_1+1}$ based on equation \eqref{eq:SplineReg} (Figure \ref{fig:first} Panel D). We use the same strategy to obtain an updated estimate $\hat{\boldsymbol{\alpha}}_{n_1+1}$, which we can update as a sample mean using the West algorithm: $\hat{Y}_{n_1+1}(t)=\hat{Y}_{n_1}(t)+\frac{I(X_{n_1+1}>t)-\hat{Y}_{n_1}(t)}{n_1+1}$. This results in an updated set of parameters $\hat{\boldsymbol{\theta}}_{n_1+1}$, and we can repeat the same process for all subsequent observations at this site, before passing the updated parameters to the next site. Every site can apply the same updating steps, until observations from all sites have been incorporated.  
The above procedure has been summarized in Algorithm \ref{algorithm:code_DKM} below. 
\spacingset{1.0} 
\begin{algorithm*}[h]
\caption{\label{algorithm:code_DKM} \small Distributed KM for updating a survival curve site-by-site.}
 \small
	\textbf{Inputs:} Survival times $\mathbf{X}=\{X_1, \dots, X_N\}$ and censoring indicators $\mathbf{\Delta}=\{\Delta_1, ..., \Delta_N\}$ across $K$ sites with sample sizes $n_1,...,n_K$ where $N = \sum_{k=1}^K n_k$. \\
	\textbf{Outputs:} Spline parameters $\hat{\theta}_{\text{final}}$ that can be used to construct survival probability estimates.\par\smallskip
1. Use the KM method to estimate survival probabilities $\hat{S}_{n_1}(t; X, \Delta)$ at site 1.
 \par
2. Fit a logistic regression with KM estimates as outcome and time as predictor to obstain spline parameters $\hat{\theta}_{\text{old}}$ and pass $\hat{\theta}_{\text{old}}$  to site 2.
\par
3. Set $i=n_1$ and $k=2$. \par 
4. For observation $(X_{i+1}, \Delta_{i+1})$ and while $i < N$, repeat:
 \par
(a) Shift survival curve based on the updating equation:
\begin{equation*}
\hat{S}_{i+1}(t) = \hat{S}_i(t;\hat{\theta}_{\text{old}}) + \frac{\hat{\psi}(X_{i+1}, \Delta_{i+1};\hat{\theta}_{\text{old}})(t)}{i+1} 
\end{equation*} \par
(b) Obtain $\hat{\theta}_{\text{new}}$ from logistic regression using the shifted curve. Set $\hat{\theta}_{\text{old}}:=\hat{\theta}_{\text{new}}$ and $i:=i+1$. \par
(c) If no observations remain at site $k$, pass $\theta_{\text{old}}$ to site $k+1$, set $k=k+1$. \par
5. Set $\hat{\theta}_{\text{final}}:=\hat{\theta}_{\text{new}}$ 
\end{algorithm*}
 \spacingset{1.9} 

A key benefit of our method is the ability to increase the complexity of our spline approximation at later sites after more observations have been incorporated. This is in contrast to parametric and semi-parametric methods, where the number of knots or degrees of freedom needs to be fixed and can be limited by sample size constraints at the initial site.

The most accurate method for updating the curve is to update one observation at a time. However, this can be computationally intensive and unnecessary when a large number of observations have already been incorporated. We therefore developed an alternative formula for updates using multiple observations at a time (see Appendix B for details). In practice we suggest updating in batches of 5-10 observations to avoid inaccuracy and optimize speed.

\subsubsection{Weighted Distributed KM}

In observational studies of treatment effectiveness, we frequently observe imbalances in characteristics between treated and untreated individuals. A treatment might be preferentially given to patients with severe disease, or more easily accessed by individuals of higher socio-economic status. In the case of thromboembolic events following vaccination, elderly people who have a higher baseline risk of blood clots also had higher rates of COVID-19 vaccination. Adjusting for these differences in a distributed setting is challenging but crucially important. Inverse propensity weighted KM curves \citep{Xie2005} are constructed by assigning each individual a weight inversely proportional to their estimated propensity score, which is the probability of receiving their treatment assignment given their covariates. This ensures covariate balance across groups, and given the standard causal inference assumptions of exchangeability, positivity, and consistency \citep{Hernan2020}, allows us to estimate the curves that we would have observed under randomized treatment conditions. Accordingly, we adapted our method to estimate distributed IPW-KM curves based on a two round algorithm sharing only summary statistics. Consider the same setting with survival data ($\mathbf{X}, \mathbf{\Delta}$) but additionally including binary treatment indicators $\mathbf{A} = (A_1,\dots,A_N)$ and confounding variables $\mathbf{Z}=(\mathbf{Z}_1,\dots,\mathbf{Z}_N)$. To construct IPW-KM curves we start by fitting a propensity score model, defined by $g(P(A_i=1\mid \mathbf{Z_i}))=\boldsymbol{\alpha}^T\mathbf{Z_i}$, 
where $g(\cdot)$ is typically the logit link function. Then we assign weights to each individual based on their predicted probability of receiving their actual treatment assignment, $w_i = \frac{1}{\hat{p}_i}$ for treated subjects and $w_i = \frac{1}{1-\hat{p}_i}$ for untreated subjects. We propose a two-pass algorithm to estimate IPW KM curves. In the first pass we use the renewable estimation algorithm proposed by \cite{Luo2020} to fit the propensity score model. This is also a sequential distributed algorithm, which uses score and hessian aggregation to fit generalized linear models. Before the second pass the model coefficients are broadcast to each site, allowing for the calculation of an inverse propensity weight $w_i$ for each observation. For the second pass, we modify our method by using a weighted updating equation:
\begin{equation}
\hat{S}_{i+1}^{\text{IPW}}(t)=\hat{S}_{i}^{\text{IPW}}(t)+\frac{w_{i+1} \psi(X_{i+1},\Delta_{i+1})(t)}{\sum_{j=1}^{i+1}w_j}  
\end{equation}
where use our original influence function, but with updates scaled by the observation weight. We then modify our approach described in Figure \ref{fig:first} by replacing our updating equations in panel C with the weighted updating equations, or equivalently modify Algorithm 1 by replacing the updating equation in step 5a.

\subsubsection{Inference for Distributed KM}

To obtain confidence intervals for the KM estimator \eqref{eq:KM}, consider a variance estimate for the cumulative hazard (or log-survival): 
$
\widehat{\text{Var}}(\log(\hat{S}(t)))=\sum_{j:t_j\leq t} (1-d_j/n_j)^{-2}\widehat{\text{Var}}\left(d_j/n_j\right)$.
Treating the number of events $d_j$ at each time as coming from an independent binomial trial, we have $\widehat{\text{Var}}(d_j/n_j) = \frac{n_j(d_j/n_j)(1-d_j/n_j)}{n_j^2}$ and hence the result from \citep{Kalbfleisch1980}: \begin{equation}\label{eq:VARNA}
\widehat{\text{Var}}(\log(\hat{S}(t))) = \sum_{j:t_j\leq t}\frac{d_j}{n_j(n_j-d_j)}
\end{equation}
A simple application of delta method then gives rise to the exponential version of Greenwood's formula, which is used to calculate standard errors and confidence intervals for KM curves that satisfy $\hat{S}(t) \in [0,1]$:
\begin{equation}\label{eq:greenwooddisc}
\begin{split}
\hat{\text{V}}(\log(-\log\hat{\text{S}}(t)))= \frac{1}{[\log\hat{\text{S}}(t)]^2} \sum_{t_j \leq t} \frac{d_j}{n_j(n_j-d_j)}
\end{split}
\end{equation}

We directly approximate Greenwood’s formula using the estimated curve parameters at the final site with no data sharing required, using the following formula:
\begin{equation} \label{eq:greenwood}
\begin{split}
\hat{\text{V}}(\log(-\log\hat{\text{S}}(t))) = \frac{1}{N[\log\hat{\text{S}}(t)]^2} \int_{0}^{t} \frac{\hat{\lambda}(t)}{\hat{\text{Y}}(t)(1-\hat{\lambda}(t))}dt
\end{split}
\end{equation}
The details of this approximation, which is derived by considering the representation of equation \eqref{eq:greenwooddisc} in counting process notation, can be found in Appendix A. 

In clinical trials and observation studies of treatment effectiveness, the log-rank test is used for comparing the KM curves of two different groups. The log-rank test compares the number of observed events in the treated and control groups to the number of expected events under the null hypothesis of equal hazards. A large difference between observed and expected events is evidence against equal hazards. To approximate the log-rank test statistic, we construct survival curves for each group $k=1,2$ as well as the overall curve including all individuals. This allows us to approximate the log-rank test statistic using our estimated curve parameters at the final site:
\begin{equation}\label{eq:logrank}
\begin{split}
O_k &= N_k \int_0^{t_{max}} \hat{\text{Y}}_k(t) \hat{\lambda}_k(t)dt \\
E_k &= N_k \int_0^{t_{max}} \hat{\text{Y}}_k(t) \hat{\lambda}(t)dt, \enspace k=1,2. \\
\text{X}^2 &= \sum_{k=1}^2 \frac{(O_k - E_k)^2}{E_k} \sim \chi^2_{1}, \enspace k=1,2.
\end{split}
\end{equation}
In equation \eqref{eq:logrank}, $N_k$ is the sample size for group $k$, $\hat{\text{Y}}_k(t)$ is the estimated proportion of individuals remaining at-risk at time $t$ in group $k$, $\hat{\lambda}_k(t)=-\frac{\hat{S}_k^{'}(t)}{\hat{S}_k(t)}$ is the estimated hazard for group $k$ at time $t$, and $\hat{\lambda}(t)=-\frac{\hat{S}^{'}(t)}{\hat{S}(t)}$ the equivalent for the entire data, all of which are easily calculated based on our spline parameters. $O_k$ represents the observed number of events for group $k$, whereas $E_k$ represents the expected number of events under the null hypothesis of equal hazards. The derivation of this approximation can also be found in Appendix A.

\subsubsection{Inference for Weighted Distributed KM}

In the IPW KM case, variance estimation is more complicated as existing methods \citep{Xie2005} rely on the combination of individual-level survival data and individual-level weights. This can result in different variances for identical survival and at-risk curves for different sets of individual-level weights. This means that our final spline estimates do not contain enough information to compute the variance of survival probability estimates at any arbitrary time point, and that equation \eqref{eq:greenwood} is no longer valid. However, we are able to exploit a convenient property of the influence function to conduct inference at a fixed, pre-specified set of time-points. Specifically, the variance of our survival probability estimate at a particular time can be estimated by the mean value of the square of the influence function across all the observations divided by the sample size:
\begin{equation}
\begin{split}
\sqrt{N}\left(\hat{\text{S}}_{IPW}(t) - \text{S}(t)\right) \xrightarrow{d} N(0, E[w_i^2\psi(\text{S};\mathbf{X}, \mathbf{\Delta})(t)^2]).
\end{split}
\end{equation}
We apply this property to compute the variance of our survival estimates for a pre-specified set of time points, and to design a distributed weighted log-rank test statistic for group comparisons after confounder adjustment. We would note that the above formula assumes that weights are known and fixed. This is a typical assumption in  variance estimates for KM curves, even outside the distributed context, but is an opportunity for future improvements. Further details on all inference procedures are given in Appendix A.

\section{Simulations}
\label{sec:sim}

\begin{figure}[h]
\begin{center}
\includegraphics[width=4.5in]{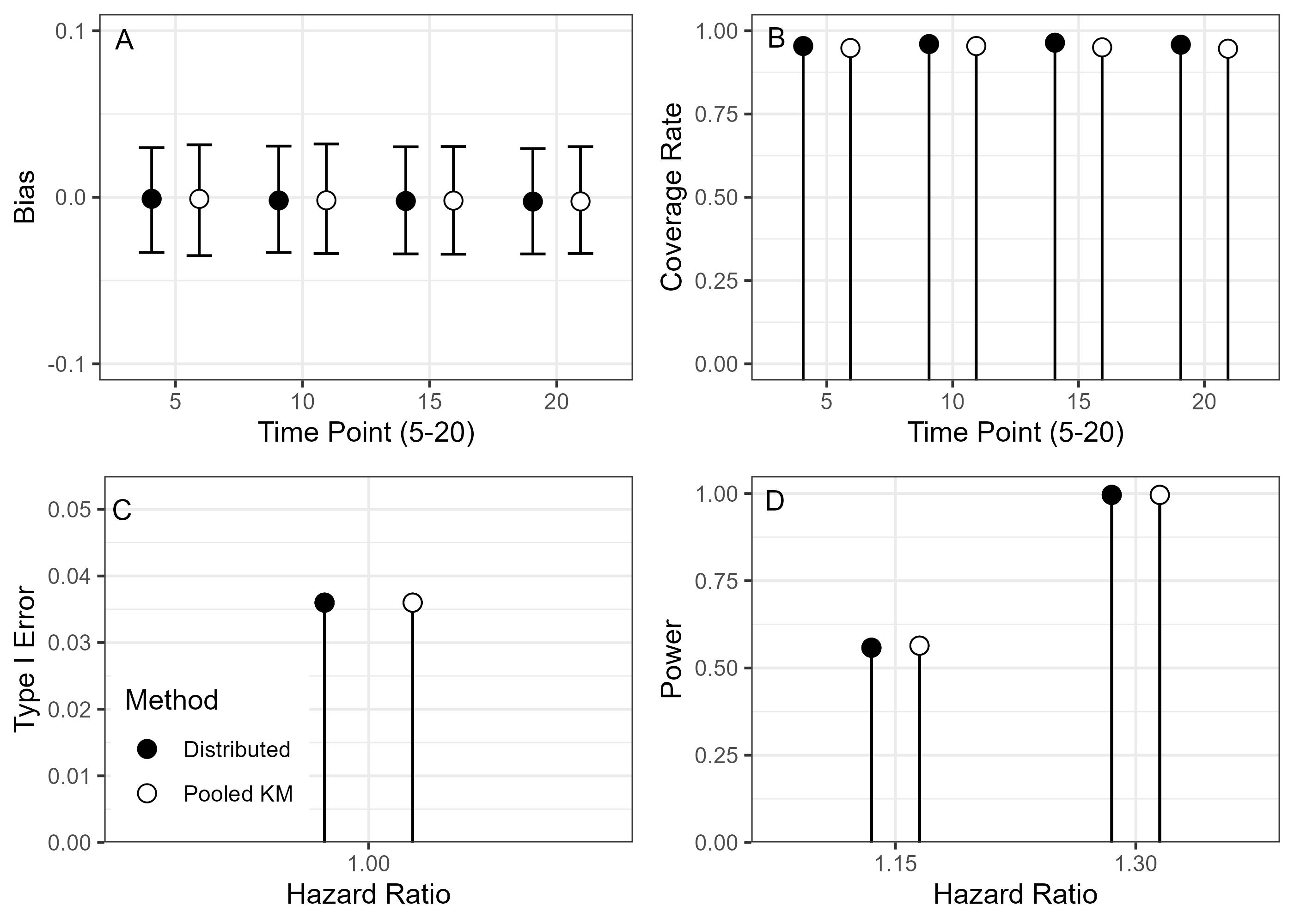}
\end{center}
\caption{Bias with empirical 95\% CI (A), empirical coverage rate (B), Type I error of log-rank test (C) and power of log-rank test (D) for simulation scenario E across 500 simulation repeats for distributed KM method compared to the oracle KM curves under complete data sharing. Simulation scenario E is described in Table 1. Abbrevs: KM; Kaplan-Meier, CI; Confidence Interval. \label{fig:second}}
\end{figure}

We evaluated our method across five simulation scenarios with varying censoring rate, hazard function, sample size, and confounding mechanism (see Table \ref{table:one} for a detailed description of all scenarios). Simulation A demonstrates the method in the simple case with no confounders, while scenarios B-E test including confounders and the robustness of our method to low sample size (B and C), more complex confounding (C-E), high censoring rate (E), and greater complexity of the survival distribution (C-E). The usage of a Weibull distribution with steep hazard (C-E) is intended to demonstrate the effectiveness of the method in one of the most difficult cases, as numerical integration can be somewhat unstable for steep hazards. 
We assessed accuracy of the estimates for survival probability at 4 time points corresponding to 25\%, 35\%, 50\% and 70\% quantiles. Results were similar across all simulation scenarios, so we present results for only the most challenging case, scenario E. Figure \ref{fig:second} shows bias and empirical coverage rate across 4 different time points, as well as type I error and power for the log-rank test with our method performing almost identically to the pooled analysis across all simulation statistics. Taken together, these metrics show that we achieve equivalent performance to the ideal analysis where patient-level data can be shared. Results for all scenarios are shown in Appendix C.

\begin{table}[ht]
\footnotesize
\captionsetup{font=footnotesize}
\begin{center}
\resizebox{\columnwidth}{!}{
\begin{tabular}{|c|c|c|c|c|c|}
\hline 
 & \multicolumn{5}{c|}{\textbf{Scenario}}\tabularnewline
\cline{2-6} \cline{3-6} \cline{4-6} \cline{5-6} \cline{6-6} 
 & \textbf{A} & \textbf{B} & \textbf{C} & \textbf{D} & \textbf{E}\tabularnewline
\hline 
\textbf{$N$ (Total)} & 800 & 800 & 800 & 2000 & 2000\tabularnewline
\hline 
\textbf{$n_{j}$ (Site)} & $5\leq n_{j}\leq350$ & $5\leq n_{j}\leq350$ & $5\leq n_{j}\leq350$ & $5\leq n_{j}\leq1000$ & $5\leq n_{j}\leq1000$\tabularnewline
\hline 
\textbf{Confounders} & None & $P(X=1)=0.5$ & \multicolumn{3}{c|}{$X_{1},X_{2}\sim N(0,1)$}\tabularnewline
\hline 
\textbf{Treatment} & $P(A=1)=0.5$ & $P(A=1\mid X)=0.3+0.4X$ & \multicolumn{3}{c|}{$P(A=1\mid X_{1},X_{2})=(1+e^{-0.5X_{1}-0.5X_{2}})^{-1}$}\tabularnewline
\hline 
\multirow{2}{*}{\textbf{Survival}} & Exponential & Exponential & \multicolumn{3}{c|}{Weibull}\tabularnewline
 & $\lambda=15$ & $\lambda=20-10X$ & \multicolumn{3}{c|}{$\lambda=12e^{0.5X_{1}+0.5X_{2}},k=0.5$}\tabularnewline
\hline 
\textbf{\% Censoring} & 30\% & 30\% & 30\% & 30\% & 50\%\tabularnewline
\hline 
\end{tabular}}
\caption{Description of simulation scenarios. All scenarios use $J=10$ sites. Survival parameters are for the control group and modified by the HR $(1.15, 1.30)$ for the treatment group in alternative scenarios. Abbrevs: HR; Hazard Ratio.\label{table:one}
}
\end{center}
\end{table}

Our method had higher coverage and slightly higher bias compared to pooled KM for most scenarios, due to some parametric smoothing in the spline approximation. This effect was relatively small and less pronounced in the high sample size scenarios (D and E) as we were able to fit splines with a greater number of knots (12 as opposed to 9). In practical we would expect overall sample sizes to be even larger, as large EHRs typically have tens of thousands to hundreds of thousands of patients.
\clearpage
\section{Application}

\subsection{Data and Study Design}

Before introducing our study design, we refer again to our main scientific questions:

\begin{enumerate}
\item	How does the risk of thromboembolic events following first COVID-19 vaccine dose compare to the risk of thromboembolic events for unvaccinated individuals with a SARS-CoV-2 infection?
\item	Does vaccination with BNT162b2 (Pfizer) or mRNA-1273 (Moderna) result in a higher risk of thromboembolic events?
\end{enumerate}

To answer these two scientific questions, we conducted a distributed analysis combining data from three large healthcare systems in Michigan: Beaumont Health, Spectrum Health, and Michigan Medicine. For the first question, we compared incidence of thromboembolic events between individuals with a SARS-CoV-2 infection (i.e. infection treated as an exposure) and individuals receiving their first COVID-19 vaccine dose (Analysis \textbf{A}). For this analysis, we categorized individuals as infected or vaccinated based on whichever occurred first, and treated the alternative exposure as a censoring event. For the second question we conducted two analyses, comparing BNT162b2 (Pfizer) recipients  to mRNA-1273 (Moderna) recipients after the first dose (Analysis \textbf{B}) and after the second dose (Analysis \textbf{C}). We used data collected between December 2020 and January 2022. Thromboembolic events were defined using ICD-10 codes derived from past literature \citep{Cox2021}.

For each analysis we computed unadjusted KM curves and log-rank test, as well as IPW curves and log-rank test to balance age, gender, race, Charlson Comorbidity index (CCI), and number of prior visits. We excluded subjects with a thromboembolic event in one year prior to baseline (vaccination or infection). Patients were subject to censoring based on death, an additional exposure (vaccine dose or infection), or study end date.

\subsection{Results}

\subsubsection{Analysis A: Infection compared to First Dose Vaccination}

\begin{figure}
\begin{center}
\includegraphics[width=6in]{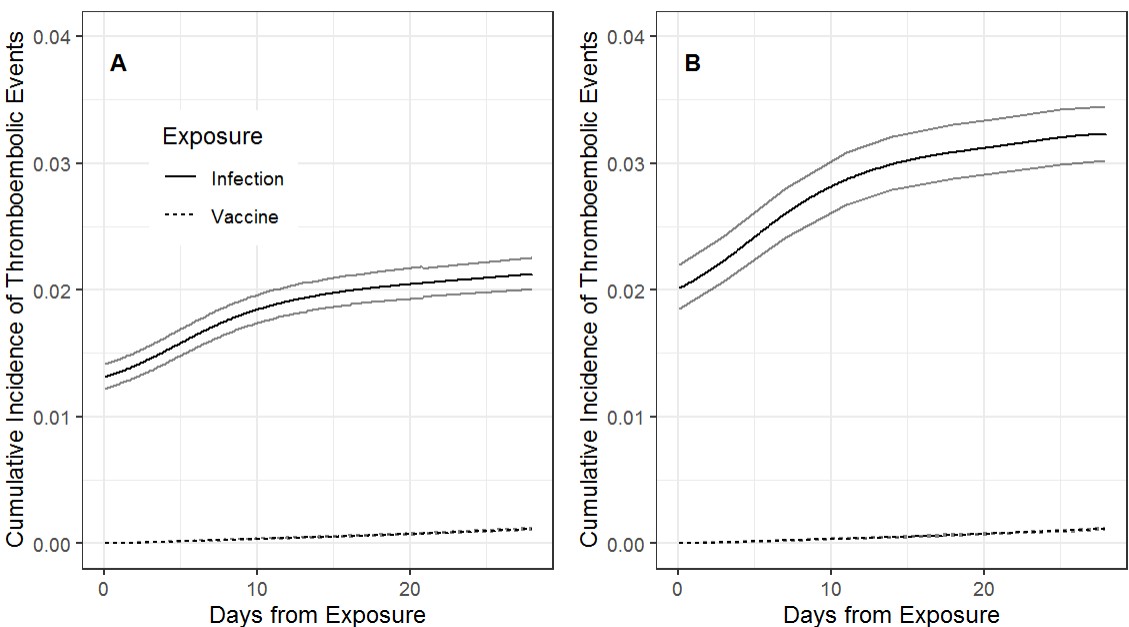}
\end{center}
\caption{Unweighted (A) and Weighted (B) Cumulative Incidence of Thromboembolic Events after First Vaccine Dose vs. after SARS-CoV-2 Infection. \label{fig:sixth}}
\end{figure}

There were $n_1=140,447$ infected or vaccinated subjects at Michigan Medicine, $n_2=478,337$ at Beaumont Health, and $n_3=353,941$ at Spectrum Health, for a total of $63,216$ infected compared to $909,509$ vaccinated subjects. Figure \ref{fig:sixth} shows unweighted (A) and weighted (B) KM curves for SARS-CoV-2 infected subjects compared to first dose recipients. In the weighted analysis, we observed a higher three-week incidence of thromboembolic events for infected (3.13\%, 95\% CI: [2.93, 3.35]) compared to vaccinated (0.08\%, 95\% CI: [0.08, 0.09]) individuals, with a significant difference in hazards based on the log-rank test ($p < 0.001$).

\subsubsection{Analysis B: First Vaccine Dose, Moderna compared to Pfizer} 

There were $n_1=140,115$ first dose recipients at Michigan Medicine, $n_2=464,911$ at Beaumont Health, and $n_3=314,122$ at Spectrum Health, with 66\% receiving BNT162b2 across all sites. In the weighted analysis, we found similar three-week incidence for mRNA-1273 (0.09\%, 95\% CI: [0.08, 0.10]) and BNT162b2 (0.08\%, 95\% CI: [0.08, 0.08]) and no significant difference in hazards based on the log-rank test ($p=0.99$). Especially given that vaccinated individuals are much older than the general population, this level of incidence is not substantially higher than what we would expect \citep{Weller2022} and should not be taken as evidence of a causal relationship.

\subsubsection{Analysis C: Second Vaccine Dose, Moderna compared to Pfizer} 

\begin{figure}
\begin{center}
\includegraphics[width=6in]{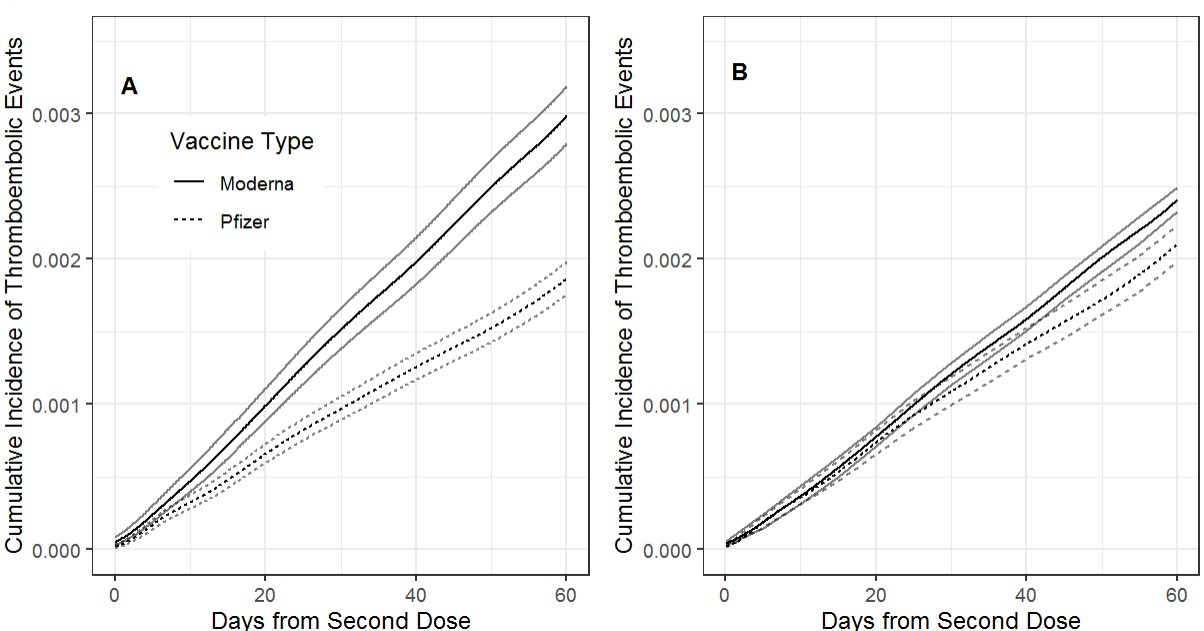}
\end{center}
\caption{Unweighted (A) and Weighted (B) Cumulative Incidence of Thromboembolic Events after Second Vaccine Dose.\label{fig:fifth}}
\end{figure}

There were $n_1=132,886$ second dose recipients at Michigan Medicine, $n_2=438,660$ at Beaumont Health, and $n_3=297,491$ at Spectrum Health, with 67\% receiving BNT162b2 across all sites. Figure \ref{fig:fifth} shows unweighted (A) and weighted (B) KM curves for BNT162b2 and mRNA-1273 following the second dose. In the weighted analysis, we observed a higher six-week incidence of thromboembolic events for mRNA-1273 (0.17\%, 95\% CI: [0.15, 0.18]) compared to BNT162b2 (0.15\%, 95\% CI: [0.14, 0.16]) recipients, with a significant difference in hazards based on the log-rank test ($p = 0.007$). As also observed in the first dose analysis, neither vaccine group had an incidence substantially higher than what we would expect given the demographic characteristics of vaccinated individuals \citep{Weller2022}.

\section{Conclusion}
\label{sec:conc}

\subsection{Multi-Center Analysis of Vaccine Adverse Events}

Using a novel method for distributed Kaplan-Meier and IPW Kaplan-Meier analysis, we were able to conduct a multi-center observational study of thromboembolic events following COVID-19 and COVID-19 vaccination across three large healthcare centers. We estimate a much higher incidence of thromboembolic events for SARS-CoV-2 infected subjects compared to vaccine recipients, in line with other findings suggesting that the benefit of vaccines in preventing COVID-19-related adverse events outweighs the risk of vaccine-related adverse events \citep{Cox2021, Tran2024}. Like all analysis using electronic health records data, a limitation of the infection analysis is that our incidence estimates are affected by ascertainment bias; patients with a thromboembolic event or other severe complications are highly likely to be tested on arrival in hospital. In contrast, mild or asymptomatic cases that do not result in adverse outcomes often go unreported. This means that we may overstate the incidence of blood clots in the infected population. Despite these limitations, we believe that our analysis demonstrates the huge potential for distributed analysis to provide high quality evidence to support time-sensitive regulatory decision-making. 

\subsection{Distributed Survival Analysis}

Distributed survival analysis has many potential applications beyond adverse event monitoring. Many treatments are targeted at rare conditions (e.g. some auto-immune diseases and cancers), and so distributed analysis is required where no single center has sufficient sample size to assess efficacy or safety. We fill a major gap in research on distributed survival analysis by presenting a method for constructing Kaplan-Meier and IPW Kaplan-Meier curves based on data from multiple sites without sharing patient-level characteristics or exact survival times. Our method offers a non-parametric alternative to existing semi-parametric and parametric methods \citep{Duan2020, Li2022} with the same desirable privacy-preserving and communication-efficient properties. Although parametric and semi-parametric methods are attractive for some applications, KM curves are a crucial tool for visualization and offer superior interpretability without parametric assumptions. Compared to differential privacy and encryption methods for non-parametric survival analysis \citep{Froelicher2022, Vogelsang2020}, we avoid administrative overhead from the involvement of third party applications or complex encryption and noise-adding procedures. We also develop a comprehensive set of methods for constructing confidence intervals and performing statistical inference. We allow for incorporation of small sites with as little as one observation, where meta-analysis or methods that require convergence of local models often fail. We provide an easy-to-use R package for practitioners, with detailed instructions for installation and use in Appendix B.

Although our method requires some parametric smoothing, an attractive feature of the updating step is that new knots can be added and spline degrees of freedom can be increased at later sites without compromising inference or estimation. This means that we can fit a highly flexible spline even when the first site is only of moderate sample size as long as the total sample size is relatively large. This is not the case for parametric estimation methods that rely on aggregated score and hessian matrices. As with many other distributed learning methods, an assumption of our method is that data is homogeneous across sites which is likely to be violated to some degree in practical applications.

The influence function has been used in other applications such as targeted maximum likelihood estimation and interpretable machine learning \citep{Laan2006, Koh2017}, but we believe that our approach of using the influence function to conduct distributed learning is entirely novel and opens new avenues for survival analysis. For example, it can be easily extended to perform distributed Cox regression \citep{Reid1985} and adapted to survival models with competing risks. It also has the potential to be applied to more complex survival models, such as distributed restricted mean survival models and multi-state survival models.

\bigskip
\begin{center}
{\large\bf SUPPLEMENTARY MATERIAL}
\end{center}

\begin{description}

\item[Supplement:] Technical appendix including theoretical details (A), github link to R packages along with detailed user instructions (B), and additional simulation results (C). (pdf)

\end{description}

\section{Appendix A: Theoretical Results}

\subsection{The Influence Function} 

To describe the influence function of the KM estimator, we start by defining the generating distribution of the censored data pairs $(X, \Delta)$ in terms of the cumulative subdistribution function $G$:
\begin{equation*}
\begin{split}
G(t, \delta)=\mathbb{P}(X < t, \Delta = \delta)
\end{split}
\end{equation*}
The true survival function $S$ admits representation as a functional $\Psi(G)$:
\begin{equation*}
\begin{split}
S(t) = \Psi(G)(t) = \exp\left(-\int_0^{t}\frac{dG(s,0)}{G(s,0)+G(s,1)}\right)
\end{split}
\end{equation*}
Peterson showed in 1977 that the KM estimate $\hat{S}$ is the result of applying $\Psi$ to the empirical cumulative subdistribution function of the data $\hat{G}$: 
\begin{equation*}
\begin{split}
\hat{S}(t) = \Psi(\hat{G})(t) = \exp\left(-\int_0^{t}\frac{d\hat{G}(s,0)}{\hat{G}(s,0)+\hat{G}(s,1)}\right)
\end{split}
\end{equation*}
To compute the influence function, we apply a small perturbation to G, based on a distribution placing all of it's mass at a single observation $(X_i, \Delta_i)$:
\begin{equation*}
\begin{split}
G_{\epsilon}(t, \delta) = (1-\epsilon)\mathbb{P}(X < t, \Delta = \delta) + \epsilon I(\Delta_i = \delta)I(X_i < t)
\end{split}
\end{equation*}
Then the influence function of $S$ is given by:
\begin{equation*}
\begin{split}
\psi(X_i, \Delta_i)(t) = \left.\frac{\partial \Psi(G_{\epsilon})(t)}{\partial \epsilon}\right|_{\epsilon=0}
\end{split}
\end{equation*}

\subsection{Update Equation}

The Von Mises expansion (Gill, 1989) of the Kaplan-Meier functional can be written as: 
\begin{equation*}
\begin{split}
S_N(t) = S(t) + \frac{1}{N}\sum_{i=1}^N\psi(S;X_i, \Delta_i)(t)+ o_p\left(\frac{1}{N^{1/2}}\right)
\end{split}
\end{equation*}
Where $\psi$ is the influence function of the KM functional. By subtracting the Von Mises expansion of  $\hat{S}_N(t)$ from the corresponding expansion for $\hat{S}_{N+1}(t)$, we can write:
\begin{equation*}
\begin{split}
\hat{S}_{N+1}(t) - \hat{S}_N(t) &= \frac{1}{N+1}\sum_{i=1}^{N+1}\psi(S;X_i, \Delta_i)(t) \\&- \frac{1}{N}\sum_{i=1}^N\psi(S;X_i, \Delta_i)(t) \\& + o_p\left(\frac{1}{N^{1/2}}\right)
\end{split}
\end{equation*}
For ease of notation we set $\psi_i = \psi(S;X_i, \Delta_i)(t)$. Moving the $\hat{S}(t)^{(n)}$ to the right side and evaluating the sum we have:
\begin{equation*}
\begin{split}
&\hat{S}_{N+1}(t)\\&= \hat{S}_N(t) + \frac{\psi_{N+1}}{N+1} - \frac{\sum_{i=1}^N\psi_i}{N(N+1)}  + o_p\left(\frac{1}{N^{1/2}}\right) \\&= \hat{S}_N(t) + \frac{\psi_{N+1}}{N+1} + o_p\left(\frac{1}{N^{1/2}}\right)\\&= \hat{S}_N(t) + \frac{\psi(S;X_{N+1}, \Delta_{N+1})(t)}{N+1} + o_p\left(\frac{1}{N^{1/2}}\right)
\end{split}
\end{equation*}
where the third term of the second line above is absorbed into the error term as $n^{-1/2}\sum_{i=1}^n\psi_i$ is $o_p(1)$ (see James 1997). \par  

\subsection{Estimation of Influence Function}

To shift the curve with new data using the influence function we need to estimate three unknown components: ${S}(t), {Y}(X_k)$ and the integral term $\int_0^{X_k \land t} \frac{\hat{\lambda}(u)}{\hat{Y}(u)}du$. Given our spline parameters $\hat{\boldsymbol{\theta}}=(\hat{\boldsymbol{\beta}}=(\hat{\boldsymbol{\beta}}_S, \hat{\boldsymbol{\beta}}_Y), \textbf{R}, p)$, we can estimate the hazard, at-risk, and survival function as follows:
\begin{equation*}
\begin{split}
\hat{S}(t; \boldsymbol{\theta})&=g^{-1}\left(\hat{\boldsymbol{\beta}}_S^T\textbf{I}(t;\textbf{R}, p)\right)\\
\hat{Y}(t; \boldsymbol{\theta})&=g^{-1}\left(\hat{\boldsymbol{\beta}}_Y^T\textbf{I}(t;\textbf{R}, p)\right) \\
\hat{\lambda}(t; \boldsymbol{\theta})&= -\frac{\hat{S}(t; \boldsymbol{\theta})}{\partial\hat{S}(t; \boldsymbol{\theta})/\partial t}
\end{split}
\end{equation*}
Here the third equation is easily computed for any standard choice of differentiable spline functions. Then the integral term $\int_0^{X_k \land t} \frac{\hat{\lambda}(u)}{\hat{Y}(u)}du$ can be computed using standard numerical integration techniques.\par \smallskip

The estimated influence function will then be given by: 
\begin{equation*}
\begin{split}
\hat{\psi}(\hat{S}; X_k, \Delta_k)(t) = -\hat{S}(t; \boldsymbol{\theta})\left[\frac{\Delta_kI(X_k \leq t)}{\hat{Y}(X_k; \boldsymbol{\theta})} - \int_0^{X_k \land t} \frac{\hat{\lambda}(u; \boldsymbol{\theta})}{\hat{Y}(u; \boldsymbol{\theta})}du\right],
\end{split}
\end{equation*}
for observations $k=1,\dots,n$. \par \smallskip

\subsection{Variance Formula}

From (Kalbfleisch and Prentice, 1980) we have:
\begin{equation*}
\begin{split}
\hat{\text{V}}(\log(-\log\hat{\text{S}}(t))) = \frac{1}{[\log\hat{\text{S}}(t)]^2} \sum_{t_i \leq t} \frac{d_i}{n_i(n_i-d_i)}
\end{split}
\end{equation*}
In counting process notation we can write this as follows, where $dN(t)$ is the observed event count at $t$ and $R(t)$ is the size of the risk set:
\begin{equation*}
\begin{split}
\hat{\text{V}}(\log(-\log\hat{\text{S}}(t))) &= \frac{1}{[\log\hat{\text{S}}(t)]^2} \int_0^t\frac{dN(u)}{R(u)(R(u)-dN(u))}du \\ &= \frac{1}{[\log\hat{\text{S}}(t)]^2} \int_0^t\frac{\hat{\lambda}(u)}{R(u)(1-\hat{\lambda}(u))}du\\ &= \frac{1}{N[\log\hat{\text{S}}(t)]^2} \int_{0}^{t} \frac{\hat{\lambda}(t)}{\hat{\text{Y}}(t)(1-\hat{\lambda}(t))}dt
\end{split}
\end{equation*}
which corresponds exactly to the variance formula quoted in the text. Above we have noted that $\frac{dN(t)}{R(t)}$ is an estimate for the hazard at $t$ and the estimate of the risk set function can be expressed as $\hat{Y}(t) = \frac{R(t)}{N}$. \par

\subsection{Unweighted Log-Rank Test}

We start from the following form of the log-rank test:
\begin{equation*}
\begin{split}
\text{X}^2 = \sum_{i=1}^2 \frac{(\sum_{t_j}O_{ij} - \sum_{t_j}E_{ij})^2}{\sum_{t_j}E_{ij}} \sim \chi^2_{1}
\end{split}
\end{equation*}
and we again reach our formulae via counting process notation:
\begin{equation*}
\begin{split}
\sum_{t_j}O_{ij} &= \sum_{t_j}N_{ij} \frac{O_{ij}}{N_{ij}} \\
&= \int_{0}^{\infty} R_i(t)\frac{dN_i(t)}{R_i(t)}dt\\
&= N_i\int_{0}^{\infty} \hat{Y}_i(t)\hat{\lambda}_i(t)dt\\
\sum_{t_j}E_{ij} &= \sum_{t_j}N_{ij} \frac{O_{j}}{N_{j}} \\
&= \int_{0}^{\infty} R_i(t)\frac{dN(t)}{R(t)}dt\\
&= N_i\int_{0}^{\infty} \hat{Y}_i(t)\hat{\lambda}(t)dt\\
\end{split}
\end{equation*}

To conduct a log-rank test of the null hypothesis of no difference in hazards, we can then compute spline-based estimates of the observed and expected number of events for each group $i$:
\begin{equation*}
\begin{split}
O_i &= N_i \int_0^{t_{max}} \hat{\text{Y}}_i(t) \hat{\lambda}_i(t)dt \\
E_i &= N_i \int_0^{t_{max}} \hat{\text{Y}}_i(t) \hat{\lambda}(t)dt, \enspace i=1,2. \\
\end{split}
\end{equation*}
where $\hat{\lambda}_i$, $\hat{Y}_i$ are the estimated hazard and at-risk functions in group $i$ and $\hat{\lambda}$ is the hazard function in the full sample (both groups). Then the log-rank statistic is given by:
\begin{equation*}
\begin{split}
\text{X}^2 = \sum_{i=1}^2 \frac{(O_i - E_i)^2}{E_i} \sim \chi^2_{1}, \enspace i=1,2.
\end{split}
\end{equation*}

\subsection{Inference for Weighted KM}

To avoid the complex and possibly intractable problem of calculating this variance without sharing subject-level data we note the following property of the influence function \cite{Reid1981}:
\begin{equation*}
\begin{split}
\sqrt{n}\left(\hat{\text{S}}(t) - \text{S}(t)\right) \xrightarrow{d} N(0, E[\psi(\text{S};X, \Delta)(t)^2]).
\end{split}
\end{equation*}
The weighted estimator has a similar property: 
\begin{equation*}
\begin{split}
\sqrt{n}\left(\hat{\text{S}}_w(t) - \text{S}_w(t)\right) \xrightarrow{d} N\left(0, E\left[\frac{w^2}{E(w)^2}\psi(\text{S}_w;X, \Delta)(t)^2\right]\right).
\end{split}
\end{equation*}
Given normalized weights, this induces the following weighted estimator of the variance of $\hat{\text{S}}_w$:
\begin{equation*}
\begin{split}
\widehat{\text{Var}}(\hat{\text{S}}_w(t)) = \frac{1}{N^2} \sum_{i=1}^N w_i^2[\hat{\psi}(\text{S}_w; X_i, \Delta_i)(t)]^2.
\end{split}
\end{equation*}
This allows us to calculate confidence intervals for a pre-specified set of time points (e.g. 6, 12, and 18-month survival) by updating and passing the cumulative sum above across each of the sites.\par \smallskip
As might be expected, the log-rank test described in the unweighted case is not valid in the weighted case because it does not account for increased uncertainty from unequal weighting of the observations. Consider the following statistic, which is the numerator of a version of the log-rank test normalized by sample size:
\begin{equation*}
\begin{split}
L &= \frac{O_1 - E_1}{N} \\
O_i &= N_i \int_0^{t_{max}} \hat{\text{Y}}_i(t) \hat{\lambda}_i(t)dt \\
E_i &= N_i \int_0^{t_{max}} \hat{\text{Y}}_i(t) \hat{\lambda}(t)dt, \enspace i=1,2 . 
\end{split}
\end{equation*}
We can estimate the variance of $L$, again using a cumulative sum:
\begin{equation*}
\begin{split}
\widehat{\text{Var}}(L) = \frac{1}{N^2} \sum_{i=1}^N w_i^2[\hat{\psi}(L; X_i, \Delta_i, A_i)]^2,
\end{split}
\end{equation*}
where the influence function of L is estimated by:
\begin{equation*}
\begin{split}
\hat{\psi}(L;X_k, \Delta_k, A_k) = &\Delta_k \left[I(A_k = 1) - \frac{\hat{Y}_{1}(X_k)\hat{p}}{\hat{Y}(X_k)} \right] \\  &- \int_0^{X_k} I(A_k = 1) \hat{\lambda}(t) - \frac{\hat{\lambda}(t)\hat{Y}_{1}(t)\hat{p}}{\hat{Y}(t)}dt,
\end{split}
\end{equation*}
for observations $k=1,\dots,n$, with $\hat{p}$ given by the total proportion of subjects who are treated. \par \smallskip
Under the null we have:
\begin{equation*}
\begin{split}
\frac{L}{\sqrt{\widehat{\text{Var}}(L)}} \xrightarrow{d} N(0,1).
\end{split}
\end{equation*}
This gives us a weighted log-rank test that we can perform without sharing subject-level data across sites. \par

\section{Appendix B: Technical Notes on Algorithm}

\subsection{Installation}

All R functions required to implement the distributed method, along with detailed documentation, can be installed using devtools::install\_github(``https://github.com/MR236/FederatedKM").\par

\subsection{Knot Selection}

We recommend selecting knot locations at equally spaced quantiles of the survival distribution as observed at the first site (as in the default settings), and using 4-10 knots with spline degree of at least 2. If the first site is large, the number of knots desired can be used at the first site. If the first site is smaller, we would recommend adding an additional knot for every 100-200 observations at subsequent sites until the target is reached. \par

\subsection{Vectorized Integration and Batch Updates} 

All the primary functions have an integration parameter (``restriction" or ``confint restriction") that controls the proportion of integration of hazards handled by Gauss-Konrod quadrature compared to Romberg integration. By default (restriction 0) we use Romberg integration for the entire follow-up, whereas setting a restriction parameter $t_r$ of greater than 0 means all integration prior to $t_r$ uses Gauss-Konrod quadrature. Typically, both methods are highly accurate and we prefer Romberg as it is much faster due to vectorization. However, in situations with very steep hazards near the beginning of the follow-up (commonly seen in adverse events data), it is less accurate. Hence we recommend setting $t_r$ to the first 5-10\% of your follow-up in all functions if you observe steep hazards. \par \bigskip

The most accurate method for updating the curve is to update one observation at a time. However, this can be computationally intensive and unnecessary when a large number of observations have already been incorporated (as the influence function estimates become very accurate). We therefore developed this alternative update formula:
\begin{equation*}
\begin{split}
\hat{S}(t)^{(n+k)} = \hat{S}(t)^{(n)} + \frac{\sum_{i=1}^k \psi(T;X_{n+i}, \Delta_{n+i})(t)}{\frac{1}{k}\sum_{i=1}^kn+i} + o_p\left(\frac{1}{n^{3/2}}\right)
\end{split}
\end{equation*}
This allows for updating $k$ observations at a time. In practice we suggest updating in batches of $0.3-0.5\%$ of the sample size already incorporated to avoid inaccuracy and optimize speed. Our R package implements batch updating for both the unweighted and weighted distributed methods. \par

\subsection{Weighted Confidence Intervals}

Due to the need to keep track of squared influence function sums, it is necessary to pre-specify the desired time points to compute weighted confidence intervals. In practice, this represents no major issue as studies typically report incidence or survival at only landmark times (e.g. 3 weeks, 6 weeks etc). In the application for this paper, we have picked an extremely dense set of time points (every 4-6 days) to demonstrate that the method can produce essentially continuous weighted confidence bands if necessary (see Figure 1B and 2B). This is computationally intensive, but feasible. It took approximately eight hours total computation time per analysis across sites with $N\approx 900,000$ and 20 time points. In contrast, 1-2 time points took 30-40 minutes. \par \bigskip

\cleardoublepage

\section{Appendix C: Additional Simulation Results}

\begin{figure}[H]
\begin{center}
\includegraphics[width=4.5in]{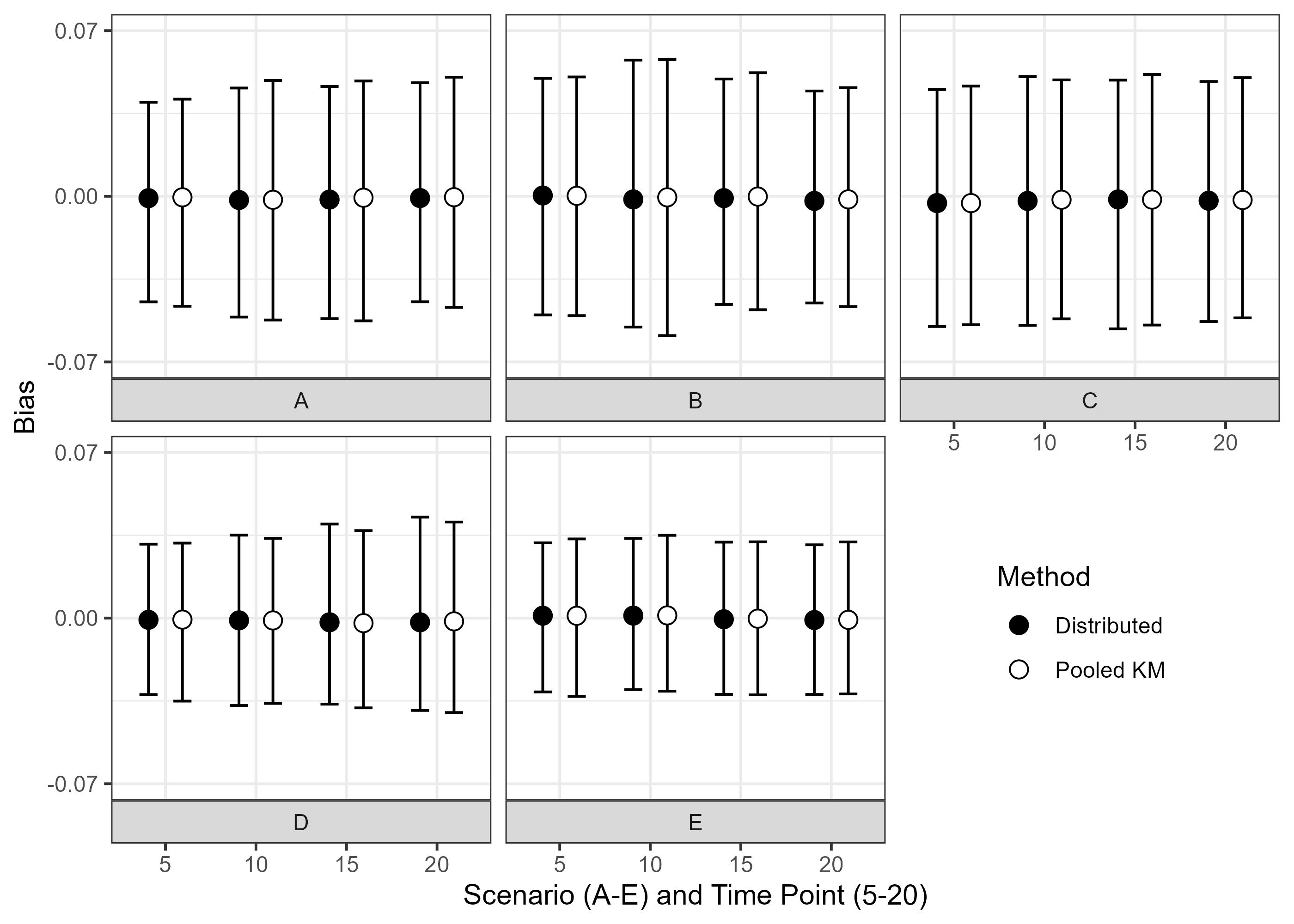}
\end{center}
\caption*{\textbf{Figure S1:} Bias with empirical 95\% CI across 500 simulation repeats for distributed KM method compared to the oracle KM curves under complete data sharing. Simulation scenarios A-E are described in Table 1. Abbrevs: KM; Kaplan-Meier, CI; Confidence Interval. \label{fig:S1}}
\end{figure}

\begin{figure}[H]
\begin{center}
\includegraphics[width=4.5in]{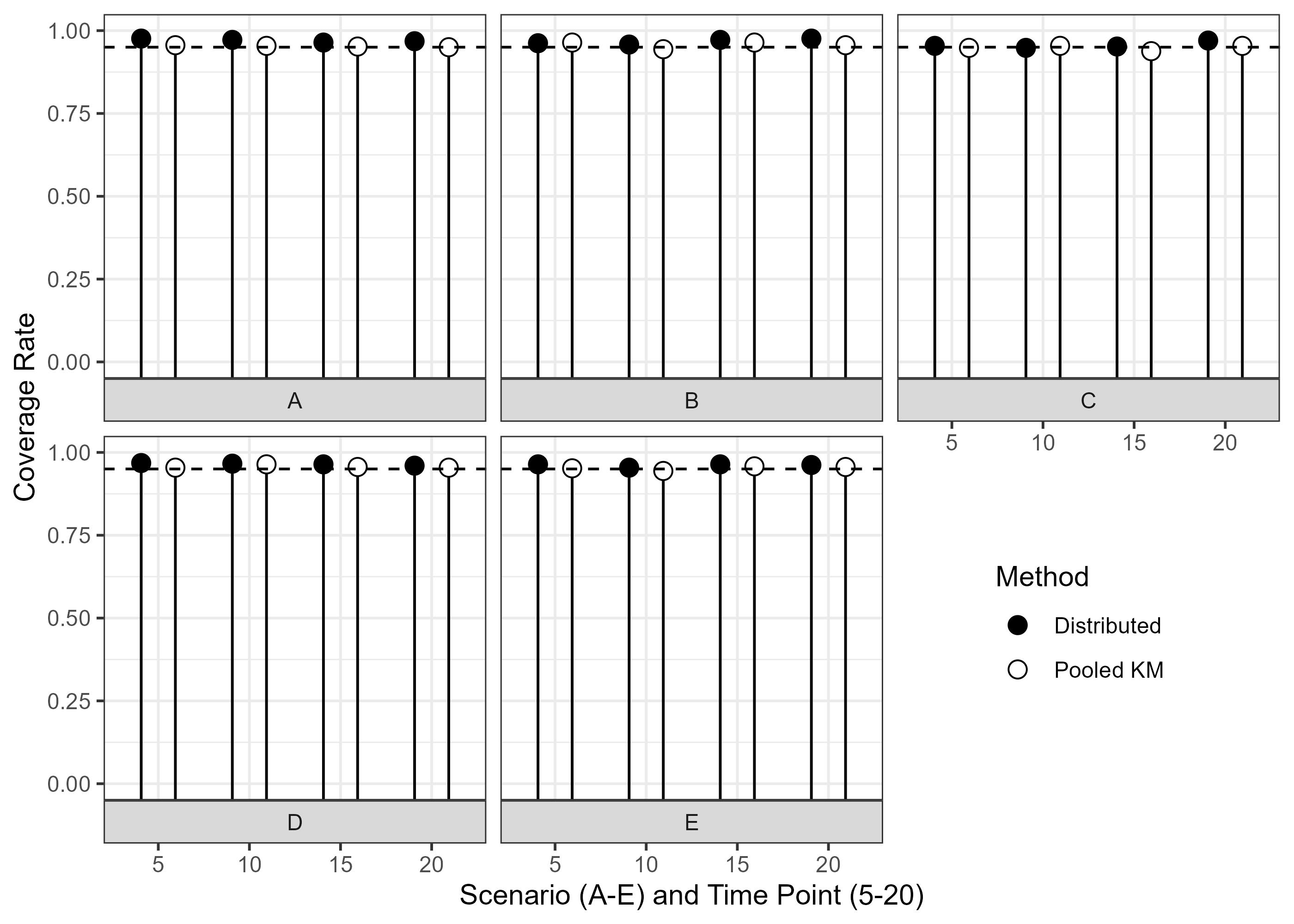}
\end{center}
\caption*{\textbf{Figure S2:} Empirical coverage rate across 500 simulation repeats for distributed KM method compared to the oracle KM curves under complete data sharing. Simulation scenarios A-E are described in Table 1. Abbrevs: KM; Kaplan-Meier. \label{fig:S2}}
\end{figure}

\begin{figure}[H]
\begin{center}
\includegraphics[width=4.5in]{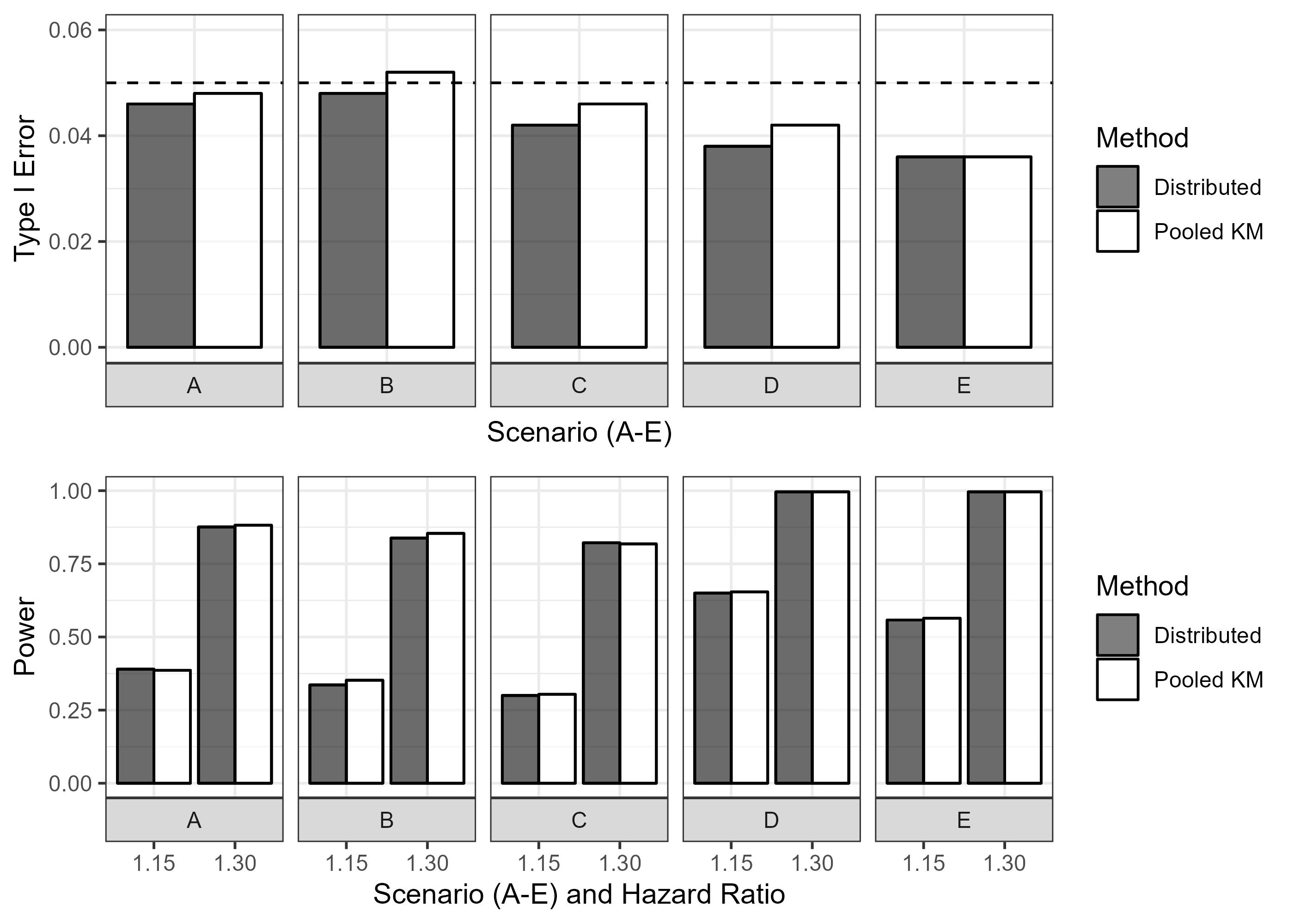}
\end{center}
\caption*{\textbf{Figure S3:} Type I error (top) and power (bottom) across 500 simulation repeats for distributed log-rank test compared to the ideal analysis (log-rank test under complete data sharing). Simulation scenarios A-E are described in Table 1. Abbrevs: KM; Kaplan-Meier. \label{fig:S3}}
\end{figure}

\section{References}

Hines O., Dukes O., Diaz-Ordaz K. and Vansteelandt S. (2022). Demystifying Statistical Learning Based on Efficient Influence Functions. \textit{The American Statistician}, \textbf{76(3),} 292-304. \bigskip

Gill, R. D., Wellner, J. A., and Præstgaard, J. (1989). Non- and Semi-Parametric Maximum Likelihood Estimators and the Von Mises Method (Part 1) [with Discussion and Reply]. \textit{Scandinavian Journal of Statistics}, \textbf{16(2),} 97–128.\bigskip

James, L. F. (1997). A study of a class of weighted bootstraps for censored data. \textit{Annals of Statistics}, \textbf{25,} 1595–1621.\bigskip

Peterson, A.V. (1977). Expressing the Kaplan-Meier estimator as a function of empirical sub-survival functions. \textit{JASA}, \textbf{72,} 854-858.\bigskip

\end{document}